\documentclass[aps,pra,twocolumn,superscriptaddress,nofootinbib,showpacs]{revtex4-1}

\usepackage{amsfonts}

\usepackage{graphicx}
\usepackage{amsmath}
\usepackage[sort&compress]{natbib}
\usepackage{hyperref}
\usepackage{braket}
\usepackage{overpic}
\usepackage{tikz}
\usetikzlibrary{matrix}

\usepackage[thickqspace,squaren]{SIunits} 

\newsavebox{\measurebox}

\begin{document}

\title{Control of multiple excited image states around segmented carbon nanotubes}

\author{J. Kn\"orzer}
\email[e-mail:\,]{johannes.knoerzer@physnet.uni-hamburg.de}
\affiliation{Zentrum f\"{u}r Optische Quantentechnologien, Universit\"{a}t
Hamburg, Luruper Chaussee 149, 22761 Hamburg, Germany}
\author{C. Fey}
\email[e-mail:\,]{christian.fey@physnet.uni-hamburg.de}
\affiliation{Zentrum f\"{u}r Optische Quantentechnologien, Universit\"{a}t
Hamburg, Luruper Chaussee 149, 22761 Hamburg, Germany}
\author{H. R. Sadeghpour}
\affiliation{ITAMP, Harvard-Smithsonian Center for Astrophysics, 60 Garden Street, Cambridge, MA 02138, USA}
\author{P. Schmelcher}
\affiliation{Zentrum f\"{u}r Optische Quantentechnologien, Universit\"{a}t
Hamburg, Luruper Chaussee 149, 22761 Hamburg, Germany}
\affiliation{The Hamburg Centre for Ultrafast Imaging,
Luruper Chaussee 149,
22761 Hamburg, Germany}
       
\date{\today}

\begin{abstract}
Electronic image states around segmented carbon nanotubes can be confined and shaped
along the nanotube axis by engineering the image potential. We show how several such
image states can be prepared simultaneously along the same nanotube. The inter-electronic
distance can be controlled \textit{a priori} by engineering tubes of specific geometries.
High sensitivity to external electric and magnetic fields can be exploited to manipulate
these states and their mutual long-range interactions.
These building blocks provide access to a new kind of tailored interacting quantum systems.
\end{abstract}


\maketitle

\section{Introduction}
Image-potential states above metallic surfaces that arise from the interplay
between an induced attractive image potential and a repulsive surface barrier
have been of considerable interest in surface studies of conducting materials
\cite{echenique78,echenique90,echenique02,kern10}. Experimentally, these
states can be observed with the aid of high-resolution
time-resolved spectroscopy techniques \cite{hoefer97} and their lifetimes were
found to be on the order of tens of ps \cite{echenique_lf_01,echenique_lf_02}.
The energy levels of such states form a Rydberg series in the Coulomb-like
image potential
$V(z) \approx - [0.85 e^2 (\varepsilon - 1)] / [(\varepsilon + 1)4z]$ \cite{cole69}
in a distance $z$ above the flat surface with dielectric constant $\varepsilon$.

In contrast, electronic states equipped with a non-zero angular momentum
$l$ around infinitely long nanowires such as \textit{carbon nanotubes} (CNTs) \cite{tis},
or fullerenes \cite{buckyballs}, stem from the balance between image potential
and centrifugal motion around nanostructure and are thus much more detached from the
metallic surface. \textit{Tubular image states} (TIS) surrounding CNTs were predicted some
years ago, \cite{tis} and experimentally observed shortly afterward \cite{zamkov_ex}.
High-angular momentum states with $\ell \gtrsim 5$ are fairly detached from the nanotube's
surface and are predicted to possess comparatively long lifetimes of
$\tau \approx \unit{1}{\nano\second} - \unit{1}{\micro\second}$ \cite{segal05},
but even low-$\ell$ TIS \cite{zamkov_th} have been shown to decay considerably slower than
their counterparts above flat surfaces \cite{zamkov_ex}.
Extensive work has been done on Bloch states in lattices of CNTs aligned parallel \cite{nanolattices}
and periodically arranged finite segments combined by junctions constituting a single, long CNT \cite{segal04}.
Furthermore, image-potential states in other systems \cite{platzman_liquidhelium,dykman_liquidhelium} have
been considered in the context of quantum information processing. Since they
are long-lived states and highly controllable by external fields \cite{segal_tuning,segal_chaos}, Rydberg-like TISs
turn out to be interesting candidates for such purposes.

Our aim is to propose a scheme for the simultaneous preparation of multiple excited
states around CNTs, allowing for subsequent studies of the long-range interactions in those systems.
We consider single CNTs whose electronic properties give rise to confinement potentials along the longitudinal
axis of the nanotube. Thereby, two external charges outside the CNT can be held at a controllable distance.
We have found that specifically engineered CNTs in the presence of an external charge give rise to highly
adjustable image potentials.

The present work is organized as follows:
in Sec.~\ref{sec:ips} we describe the general properties of tubular image states around infinitely long nanowires.
We address the question of how they are controllable in Sec.~\ref{ssec:bfields}.
Subsequently, in Sec.~\ref{ssec:inf_to_fin}, the properties of finite and infinite geometries are interlinked.
Sec.~\ref{ssec:trapping} is dedicated to a general description of trapping two or more electrons in tubular image
states around one segmented CNT, where the inter-electronic distance is, in principle, \textit{a priori}
controllable by geometry.
Furthermore, in Sec.~\ref{ssec:design} we investigate segmented arrays \cite{segal04} of nanotubes and show that by
considering also not perfectly conducting CNTs \cite{arista12}, in principle arbitrary potentials can be
generated along the longitudinal axis of the nanotube.
As an outlook, we sketch potential applications and the description of long-range interactions between two image
states in Sec.~\ref{sec:outlook}.

\section{Cylindrical image states above nanowires}\label{sec:ips}

As for charged particles in the vicinity of flat metallic surfaces, the method of
images \cite{jackson} can be applied in order to describe the attractive force
between an infinitely-long hollow metallic cylinder of radius $a$ and a charge $q$
at $\mathbf{r}_0 = (\rho_0>a, \ z_0, \ \varphi_0)$.
The polarization of the cylindrical surface can be described by the induced
scalar potential $\Phi_{\text{ind}}$ at $\mathbf{r}=(\rho, \ z, \ \varphi)$
via an expansion in terms of regular and irregular Bessel functions \cite{arista01}, $K_\alpha$ and $I_\alpha$, $\alpha \in \mathbb{Z}$, respectively:

\begin{eqnarray}\label{eq:potind}
\Phi_{\text{ind}}(\mathbf{r},\mathbf{r}_0) = - \frac{2q}{\pi} \underset{\alpha=-\infty}{\overset{\infty}{\sum}} \underset{0}{\overset{\infty}{\int}} d k \cos [ k (z-z_0) ] \nonumber \\
\times \exp [ i \alpha (\varphi-\varphi_0) ] \frac{I_{\alpha}(ka)}{K_{\alpha}(ka)} K_{\alpha}(k\rho_0) K_{\alpha}(k \rho).
\end{eqnarray}

Evaluated at the position of the external charge $\mathbf{r} = \mathbf{r}_0$,
this yields the image potential which the charge feels:

\begin{eqnarray}\label{eq:potim_infinite}
V_{\mbox{\tiny im}}(\rho_0) & = & \frac{q}{2} \Phi_{\mbox{\tiny ind}}\big |_{(\rho_0,0,0)} \nonumber \\
& = & -\frac{q^2}{\pi} \sum_{\alpha=-\infty}^{\infty} \underset{0}{\overset{\infty}{\int}} d k \ \frac{I_{\alpha}(ka)}{K_{\alpha}(ka)} K_{\alpha}^2(k\rho_0).
\end{eqnarray}

Asymptotic forms of this potential were derived and an analytical expression was informed
by the asymptotics \cite{tis}.
In most studies thereafter,
the analytical form was used in calculations in order to reduce numerical costs. Recently it was pointed
out \cite{arista12} that quantitative differences in eigenenergies and eigenstates existed,
using the exact, cf.\ Eq.~(\ref{eq:potim_infinite}), and approximate \cite{tis} potentials, respectively.

Introducing the repulsive angular momentum barrier, in the absence of external fields, the one-dimensional
effective potential reads

\begin{equation}\label{eq:poteff_infinite}
V_{\ell}(\rho_0) = V_{\mbox{\tiny im}}(\rho_0) + \frac{\ell^2 - 1/4}{2 m \rho_0^2},
\end{equation}

\noindent the parameter $\ell$ being the angular momentum and $m$ denoting the mass of the charged
particle, in the following assumed to be an electron. Hereafter, we use atomic units, i.e.\ $\hbar = m_e = e = (4\pi \varepsilon_0)^{-1} = 1$.
Whether or not detached bound states exist primarily depends on the angular momentum quantum number.
For too low $\ell$ the effective potential is overall attractive everywhere and for too high $\ell$ there is no
detached local potential minimum accompanying the angular momentum barrier.
In earlier works $\ell_{\mbox{\tiny min}} = 5$ was already found to be the minimum angular momentum which supports bound TIS \cite{arista12} around
infinite nanowires.

The total electronic wavefunction separates in cylindrical coordinates $(\rho,z,\varphi)$,

\begin{equation}\label{eq:states_sep}
\Psi_{\ell n k}(\rho,z,\varphi) = \frac{\psi_{\ell n}(\rho)}{\sqrt{2\pi \rho}} e^{i\ell \varphi} \phi_k(z),
\end{equation}

\noindent where $\rho = 0$ refers to the center of the tube. This allows us to express the total energy as a sum of transverse and longitudinal energies, $E_{\ell n k} = E_{\ell n} + E_k$,
where the transverse part scales for fixed $n$ as $E_{\ell n} \propto \ell^{-3}$ \cite{tis}. Typical binding energies range between
$\unit{15}{\milli\electronvolt}$ and a few meV.

The radial wavefunction $\psi_{\ell n}(\rho)$ satisfies the radial Schr\"odinger equation

\begin{equation}\label{eq:schr}
\left ( \frac{\partial^2}{\partial \rho^2} + 2 [ E_{\ell n} - V_\ell(\rho) ] \right ) \psi_{\ell n}(\rho) = 0.
\end{equation}

\noindent States with the same radial quantum number $n$, i.e.\ number of nodes of $\psi_{\ell n}$,
and different $\ell$ are non-degenerate and are,
in general, energetically farther apart than adjacent states of same $\ell$ but different $n$.
The maximum electronic density is found at a distance $\rho_{\mbox{\tiny max}}$ from the CNT,
which scales as $\propto \ell^{3}$ \cite{tis}.
This makes a certain range of angular momenta $\ell \gtrsim \ell_{\mbox{\tiny min}}$ especially interesting for
our purposes since farther away from the surface
the states are experimentally hard to protect against extraneous influence by the surrounding
medium \cite{zamkov_ex}.

\section{Control of image states}\label{sec:control}

TISs may be controlled by means of external fields. However, these do not serve as the only control parameters for the states.
Furthermore, the radius of the nanotube can be adjusted
by rolling up a graphene sheet accordingly \cite{terrones_review}.
So far, studies have focused on (10,10) armchair nanotubes ($a=\unit{0.68}{\nano\meter}$).
Tubes with smaller (larger) diameters tend to support states with smaller (greater) spatial extent. Moreover, the length and
conducting properties may alter the TIS properties as well.
\begin{figure}[t!]
\centering
 \begin{overpic}[width=0.5\textwidth]{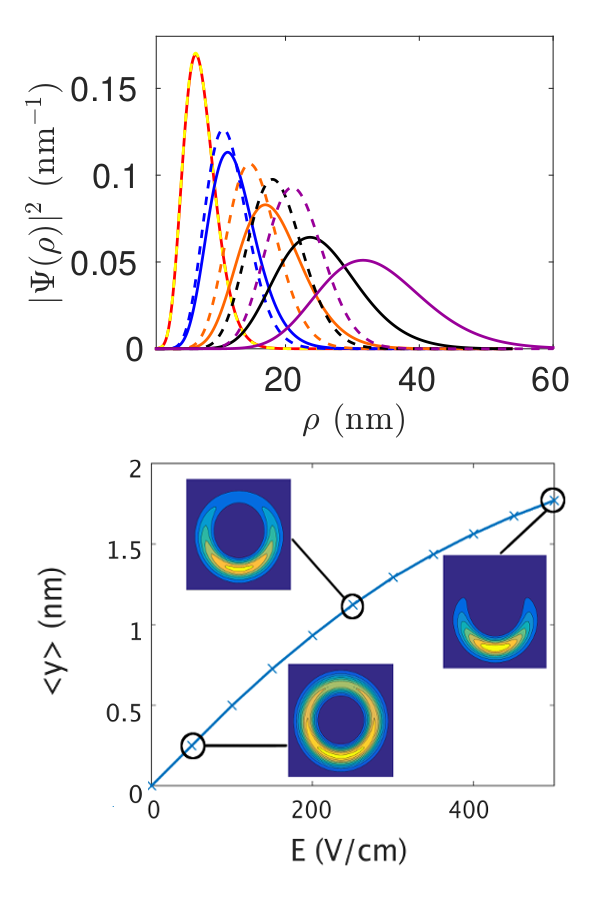}
  \put(2,95){\color{black}{\Large (a)}}
  \put(2,47){\color{black}{\Large (b)}}
 \end{overpic}
\caption{(a) Radial wavefunctions $\psi_{\ell 1}(\rho)$ for angular momenta $\ell=5-9$ \textit{(from left to right)} in the case of no external fields \textit{(solid)}
and with a magnetic field of $B=\unit{20}{\tesla}$ applied \textit{(dashed)} parallel to the longitudinal axis of the nanotube.
(b) Mean displacement $\braket{y}$ perpendicular to the tube for an electric field applied along the $y$-axis for a $(l=5, \ n=1)$-state.
The insets show the probability densities of the perturbed states for the field strengths E = 50 V/cm, 250 V/cm
and 500 V/cm, respectively. White dots indicate the nanotube.}
\label{fig:eb-states}
\end{figure}

\subsection{Impact of magnetic and electric fields}\label{ssec:bfields}

Tubular image states can be strongly affected by external magnetic fields applied parallel to the longitudinal axis $z$ \cite{segal05}.
Neglecting any contributions stemming from the electron's spin due to its global character, two additional terms, $H_Z$ and $H_d$, a Zeeman and
a diamagnetic term, respectively, enter the Hamiltonian and thus the Schr\"odinger Eq.~(\ref{eq:schr}),

\begin{eqnarray}\label{eq:zeeman_dia}
H_Z = - \frac{B}{2} L_z, \nonumber \\
H_d = \frac{B^2}{8} \rho^2,
\end{eqnarray}

\noindent where $L_z$ denotes the $z$-component of the angular momentum operator. For TISs, being highly extended, the diamagnetic
term can dominate. Higher-$\ell$ states are farther away from the tube's surface,
e.g.\ $\braket{\rho}_{\ell=5}^{n=1} \approx \unit{7}{\nano\meter}$ and $\braket{\rho}_{\ell=7}^{n=1} \approx \unit{18}{\nano\meter}$
for the $n = 1$ ground states of a (10,10) CNT of radius $a=\unit{0.68}{\nano\meter}$.
Hence, states possessing higher angular momenta are more strongly affected by the presence of a magnetic field.
The respective ground states of Eq.~(\ref{eq:schr}) are shown for $\ell = 5, ..., 9$ in Fig.~\ref{fig:eb-states}(a).
For $\ell = 5$, the states with or without an external magnetic field of $B = \unit{20}{\tesla}$ essentially coincide whereas the higher-$\ell$
states differ significantly.
Also, the spectrum is altered, again more significantly for higher-lying states.

Under the application of external electric fields perpendicular to the longitudinal axis of the nanotube, TISs lose their
rotational symmetry, as states of different $\ell$ mix.
The consequent electric-field induced decentering and distortion of the image states makes the position of maximum probability density
vary with the field strength $E$, cf. Fig.~\ref{fig:eb-states}(b). This decentering is reminiscent of a field-induced electric dipole.
Longitudinally confined states can be selectively addressed by an external electric field.
Hence, electric fields can be exploited to tune the interaction between two image states.

\subsection{Impact of finite-length segmentation}\label{ssec:inf_to_fin}

The wavefunctions in Eq.~(\ref{eq:states_sep}) reflect that the external electron is spatially not confined
along the longitudinal axis $z$.
In order to induce a confining potential along $z$, we make use of the versatile electronic properties of CNTs which,
above all, depend on the geometry of the CNT.
Finite CNT segments with different electronic properties can be merged into a single CNT of finite length which
inherits its conducting properties locally from its constituents.
Moreover, the individual segments can differ in length and diameter.
Periodic arrays of such finite-sized metallic nanotubes have been investigated with a focus on Bloch states in systems
of identical segments connected by junctions \cite{segal04}. These junctions are constructed by a rotationally asymmetric
series of pentagon-heptagon defects in the lattice structure \cite{charlier-57,chico_defects}.
In contrast, we explore CNTs consisting of non-periodically arranged segments which eventually give rise to
confinement potentials for the external electron along the axis of the nanotube.
In a next step, we develop a scheme that allows the
preparation of several such excited states around one suspended CNT, which in turn eventually
may consist of a series of metallic, semi-conducting and insulating CNT segments.

As the standard method of images in electrostatics is, for flat surfaces as for cylindrical geometries, grounded on the
assumption of an infinitely extended metal \cite{jackson},
the attractive portion of the potential \footnote{For the sake of simplicity, we will call the attractive part of the potential
also the \textit{image potential} for finite CNTs.}, binding the electron to a CNT of finite length, has to be computed differently.
Since the external electron induces a charge distribution on the nanotube surface with which it interacts, the
calculation of the image potential in the presence of a finite CNT reduces to the determination of the induced
surface charge density. A discretized charge density, consisting of $N =  10^{3}$ to $N = 10^{4}$ tile charges $\{q_i\}_{1\leq i\leq N}$ at positions
$\{\mathbf{r}_j\}_{1\leq j\leq N}$ on the surface, distributed along
and around the CNT, yields the image potential \cite{segal04} as a sum over Coulombic interactions between the external electron
and the tile charges:

\begin{equation}\label{eq:potim_finite}
V_{\mbox{\tiny im}}(\mathbf{r}_0) = - \frac{1}{2} \sum_{i=1}^{N} \frac{q_i}{|\mathbf{r}_0-\mathbf{r}_i|}.
\end{equation}

\noindent The effective potential in Eq.~(\ref{eq:poteff_infinite}) is thus the sum of the above
potential and the centrifugal barrier potential. We are now interested in the numerical solutions
of the wavefunctions $\chi_{\ell n}(\rho,z)$ from Eq.~(\ref{eq:schr2d}). Due to the
longitudinal confinement which couples $\rho$- and $z$-motion, they now depend on both $\rho$ and $z$
and are labeled by $\ell$ and $n$, a quantum number counting the energy levels. The Schr\"odinger equation,

\begin{equation}\label{eq:schr2d}
\left [ - \frac{1}{2} \left ( \frac{\partial^2}{\partial \rho^2} + \frac{\partial^2}{\partial z^2} \right )+ V_\ell(\rho, z) - E_{\ell n} \right ] \chi_{\ell n} (\rho, z) = 0,
\end{equation}

\noindent can be solved, e.g., by employing a two-dimensional finite differences method \cite{leveque_finitedifferences}
or by setting up a discrete variable representation \cite{light_dvr}.

Since the gapped excitation spectrum of longitudinal excitations features much smaller transition energies than
radial excitations, an adiabatic approximation in $z$ may be employed, reducing the numerical efforts of solving
Eq.~(\ref{eq:schr2d}) to the evaluation of two one-dimensional eigenvalue equations \cite{segal04}.
Since the primary computational cost stems from the calculation of the surface tile charges and thus the image
potential, i.e.\ Eq.~(\ref{eq:potim_finite}), whether or not the adiabatic approximation is employed is of
minor importance. Therefore,
we have in most calculations solved the full problem of Eq.~(\ref{eq:schr2d}), because especially for higher-lying
states the resulting binding energies of the exact and approximative calculations do differ up to a few percent.

\begin{table}[t!]
\begin{tikzpicture}
 \matrix (mymatrix) [matrix of nodes,
            nodes in empty cells,
text height=2.5ex,
text width=6.9ex
            ]
{
\begin{scope} \tikz\node[overlay] at (-0.2ex,0.25ex){\footnotesize $L_{\mbox{\tiny CNT}}$};\tikz\node[overlay] at (6.8ex,1.1ex){$\ell$}; \end{scope} & ~~5 & ~6 & ~7 & ~8\\\hline
\unit{0.2}{\micro\meter} & -2.8  &      &      &  \\
\unit{0.4}{\micro\meter} & -7.1  & -1.9 & -0.4 &  \\
\unit{0.8}{\micro\meter} & -10.3 & -3.9 & -1.7 & -0.7\\
\unit{1.6}{\micro\meter} & -13.5 & -6.9 & -4.2 & -2.3 \\
$\infty$ & -14.0 & -7.7 & -4.8 & -3.3\\
};
\draw (mymatrix-1-2.north west) -- (mymatrix-6-2.south west);
\draw (mymatrix-1-1.north west) -- (mymatrix-1-1.south east);

\end{tikzpicture}
\caption{Binding energies of degenerate bound states for (10,10) nanotubes of various lengths in meV. Blank cells indicate that no bound states
exist. Our $L_{\mbox{\tiny CNT}}=\infty$ results agree up to less than a percent with \cite{arista12}.}
\label{table:energies}
\end{table}

\begin{figure}[t!]
  \centering
   \begin{overpic}[width=0.5\textwidth]{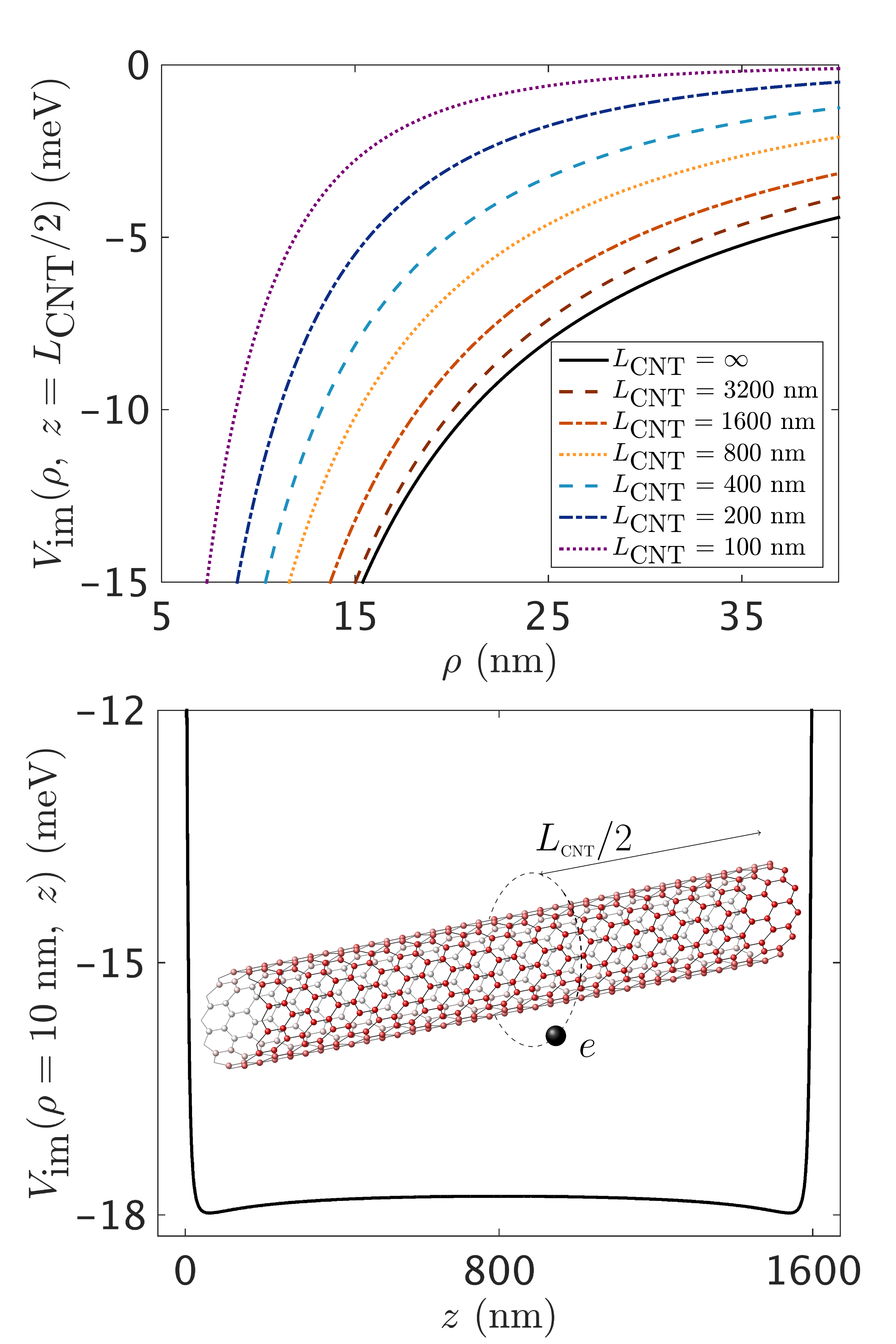}
    \put(17,90){\color{black}{\Large (a)}}
    \put(17,42){\color{black}{\Large (b)}}
    \put(21,32){\color{black}{\Large (c)}}
   \end{overpic}
\caption{
(a) Image potentials for (10,10) carbon nanotubes of various lengths. The potential is evaluated
at $z=L_{\mbox{\tiny CNT}}/2$ along the tube, where $L_{\mbox{\tiny CNT}}$ is the length
of the respective nanotube.
(b) Image potential at $\rho = \unit{10}{\nano\meter}$ and evaluated along the longitudinal axis $z$ for a
(10,10) carbon nanotube of length $L_{\mbox{\tiny CNT}} = \unit{1600}{\nano\meter}$.
(c) Sketch of CNT of length $L_{\mbox{\tiny CNT}}$. The electronic orbit at
$z=L_{\mbox{\tiny CNT}}/2$ shows where along the CNT the potential curves in (a) are evaluated.
}
\label{fig:finitecns}
\end{figure}

As to be expected, the image potential becomes more attractive as the length
of the CNT, $L_{\mbox{\tiny CNT}}$, increases, cf.\ Fig.~\ref{fig:finitecns}(a). In the depicted range,
i.e.\ $\unit{5}{\nano\meter} \leq \rho < \unit{40}{\nano\meter}$, the finite-tube results converge to the image potential
of an infinite nanowire as $L_{\mbox{\tiny CNT}}$ is increased. Of course, in the long-range limit
$\rho \rightarrow \infty$ the electron feels the difference in attraction no matter the length of the nanotube.
The lowest-lying bound states of Eq.~(\ref{eq:schr}) extend to less than $\unit{50}{\nano\meter}$ away from the
tube's surface. Hence, finite-size effects affect the image potential considerably for moderate lengths
$L_{\mbox{\tiny CNT}} \lesssim \unit{1}{\micro\meter}$. The binding energies of the resulting TIS are also
profoundly affected especially in this length regime.
This is demonstrated in Table~\ref{table:energies} for the respective $n=1$ ground states. The usual $\ell^{-3}$
scaling of the binding energies with the angular momentum is, for high-$\ell$ states, distorted by finite-size effects,
since these states are to be expected far outside and therefore witness the finite spatial extent of the CNT.
Note that finite geometries lead to edge effects that are also present here. Therefore, for the potential curves in
Fig.~\ref{fig:finitecns}(a), $V_{\mbox{\tiny im}}(\rho,z)$ is evaluated at $z=L_{\mbox{\tiny CNT}}/2$, cf.\ Fig.~\ref{fig:finitecns}(c).
Around this value, $V_{\mbox{\tiny im}}$ is symmetric,
i.e.\ $V_{\mbox{\tiny im}}(\rho,L_{\mbox{\tiny CNT}}/2-z)=V_{\mbox{\tiny im}}(\rho,L_{\mbox{\tiny CNT}}/2+z)$.
Fairly long nanotubes almost generate box potentials, with shallow wells at the ends, along their axes at radial distances
where the image-potential states
are most likely to be found, cf.\ Fig.~\ref{fig:finitecns}(b).
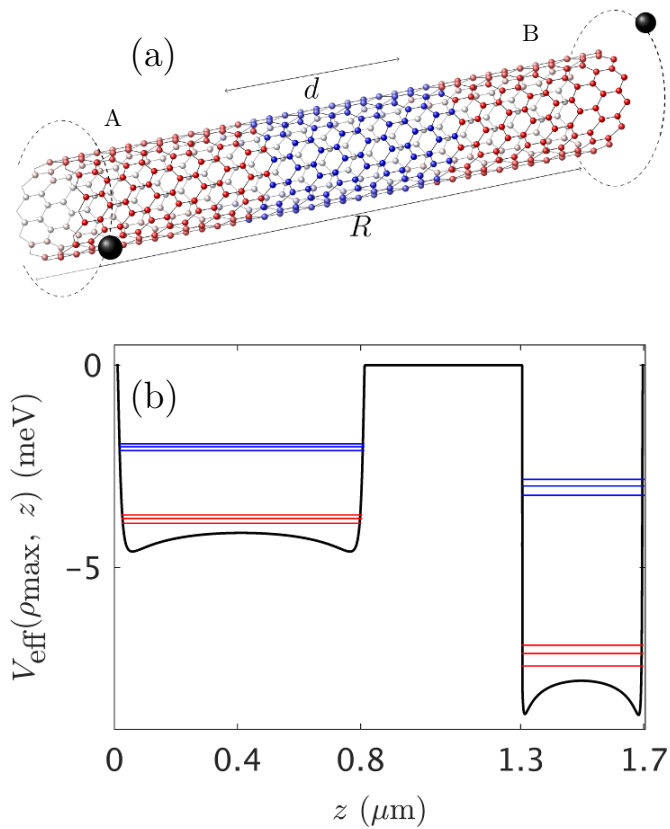
\begin{figure}[t!]
   \begin{overpic}[width=0.5\textwidth]{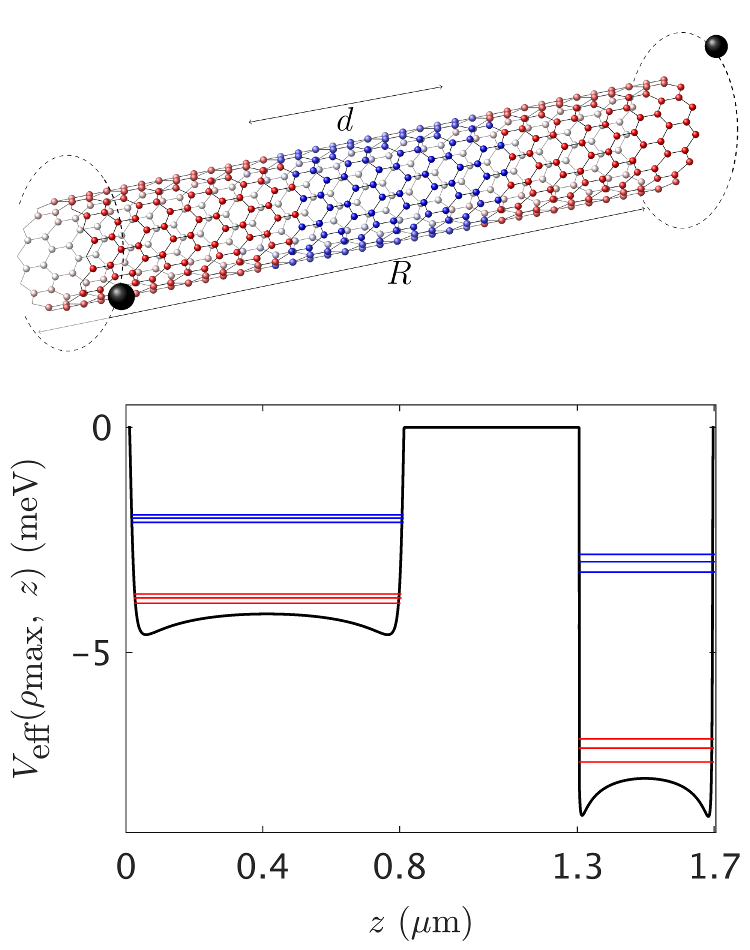}
    \put(15,90){\Large \color{black}{(a)}}
    \put(15,50){\Large \color{black}{(b)}}
    \put(12,83){\color{black}{A}}
    \put(61,93){\color{black}{B}}
   \end{overpic}
\caption{(a) Scheme of three-segment nanotube with two confined electrons in tubular image states. $d$ is the length of the insulating
segment that connects two metallic segments at whose ends the states are localized. The distance $R$ denotes the mean-value distance of
the electronic orbits.
(b) The corresponding
effective potential is evaluated along $z$ at the radial distance of maximum probability density. Energy levels of the
first three single-electron eigenstates of even parity ($n=1, \ 3, \ 5$) in the radial ground states \textit{(red)} as well as the first
three single-electron eigenstates of even parity in the first-excited \textit{(blue)} radial-state manifolds are shown for both subsystems.}
\label{fig:two_electrons}
\end{figure}
In the presence of a single segment, the energetically lowest-lying states tend to be localized at the ends of the segment.
An intuitive understanding of this can be obtained by looking at the induced charge distribution on the CNT.
This charge distribution will be symmetric about the center of the tube if the external electron is located above the
center of the CNT. Therefore, on both sides of the electron in longitudinal directions, charges drag the electron towards
the edges. The longitudinal components of these forces cancel. In contrast, if the electron is located closer to one of
the edges, the induced charge density is also located closer to this edge and as the external electron approaches the
edge, the induced charge distribution will be localized at this edge. Therefore, at fixed radial distance from the CNT,
there is a total attraction towards the edges.
A perfectly symmetric image potential around $z=L_{\mbox{\tiny CNT}}/2$ yields nearly degenerate
eigenstates of even and odd parity, respectively, which are delocalized states with significant amplitude only near the
edges, spread towards the center of the CNT segment for excited states \cite{segal04}.

\subsection{Localization of multiple excited states}\label{ssec:trapping}

An electron above a non-metallic surface does either not feel the feedback of the material in form of an induced image
potential at all, if the material is insulating, or sees a weaker image potential than in the metallic case, depending
on the precise electronic properties of the semi-conducting CNT.
Therefore, by connecting two conducting segments
by an insulator, two TISs can be spatially separated, cf.\ Fig.~\ref{fig:two_electrons}(a). In this way, the distance between
two states can be controlled \textit{a priori} via the length of the insulating segment. In Fig.~\ref{fig:two_electrons}(b), the
effective potential
$V_{\mbox{\tiny eff}}^{\mbox{\tiny AB}}(\mathbf{r}_{\mbox{\tiny A}},\mathbf{r}_{\mbox{\tiny B}}) \approx V_{\mbox{\tiny eff}}^{\mbox{\tiny A}}(\mathbf{r}_A) + V_{\mbox{\tiny eff}}^{\mbox{\tiny B}}(\mathbf{r}_B)$
as a function of $z$ at the radial distance of maximum
probability $\rho = \rho_{\mbox{\tiny max}}$ is shown for a system consisting of two spatially separated electrons both in TISs.
Along the insulating segment, the effective potential is set to zero.
The two external electrons are confined around metallic segments of
$L^{\mbox{\tiny A}}_{\mbox{\tiny CNT}} = \unit{0.8}{\micro\meter}$ and $L^{\mbox{\tiny B}}_{\mbox{\tiny CNT}} = \unit{0.4}{\micro\meter}$, respectively, with
angular momenta $\ell_{\mbox{\tiny A}} = 6$, $\ell_{\mbox{\tiny B}} = 5$.
The distance $R$ between the electronic orbits is, most of all, dictated by the length $d=\unit{0.5}{\micro\meter}$ of the
insulating and the lengths of the metallic segments, since these states have significant amplitudes only near the edges.
Also, Fig.~\ref{fig:two_electrons}(b) shows
the energy levels of the three lowest-lying even parity one-electron states in both the radial ground and first excited state, respectively. The odd and even
parity states lie energetically so close that they cannot be resolved in Fig.~\ref{fig:two_electrons}(b).
Initializing a single-electron state at one of the edges of its segment in a superposition
$\ket{L}_\sigma = 1/\sqrt{2} \ (\ket{0}_\sigma + \ket{1}_\sigma)$
or $\ket{R}_\sigma = 1/\sqrt{2} \ (\ket{0}_\sigma - \ket{1}_\sigma)$, $\sigma =$ A, B labeling the
two TISs around their segments and $\ket{0}_\sigma$ ($\ket{1}_\sigma$) denoting the corresponding even (odd) parity state,
the interaction with a second TIS can be altered significantly since thereby, the mean interelectronic
distance $R$ can be controlled.
Since the odd and even parity states lie energetically very close,
the lifetimes of the $\ket{L}_\sigma$ and $\ket{R}_\sigma$ states are comparable with the estimated lifetimes due to other decay channels.

The states can also be manipulated \textit{a priori} by inserting semi-conducting segments.
For finite dielectric constant $\varepsilon$, i.e.\ the low-frequency limit of the permittivity $\varepsilon(\omega)$,
the leading-order correction to the image potential in Eqs.~(\ref{eq:potim_infinite}) and (\ref{eq:potim_finite}) scales as $1/\varepsilon$ \cite{arista03}.
Because of the sparse literature available on dielectric
functions of CNTs in the static limit \cite{stat_limit}, a two-fluid model \cite{mowbray_twofluid} was applied to infinitely long
CNTs with the aim of finding a way to properly describe the image potentials due to semi-conducting CNTs \cite{arista12}.
However, for the sake of simplicity, we have calculated image potentials of semi-conducting CNTs in the present work by
assuming a macroscopically defined dielectric constant. For CNTs, it might be obtained from projecting the dielectric tensor of
graphite onto a cylinder \cite{stoeckli99}. The smaller the dielectric constant is, the fainter is the attractive inward force on
the electron due to the image potential.
It has been remarked \cite{arista12} that the band gap of a semi-conducting CNT should not exceed a value of $\unit{0.1}{\electronvolt}$ in order to support bound states
in the effective potential.
Ultimately, the set of techniques for tuning TIS mentioned so far allows for preparing the electrons in specific states along a CNT
and, in the presence of external fields, with specific permanent dipole moments which can lead to controlled long-range
interactions between two ore more image states \cite{saffman_rydbergreview}.

\subsection{Engineering of image potentials}\label{ssec:design}

The electronic properties of carbon nanotubes are determined by their chiral indices $(n,m)$ \cite{saito98,charlier07}.
Therefore, the dielectric constants vary from nanotube to nanotube.
Hence, by merging different CNTs, step-wise potentials can be generated along the longitudinal axis of a segmented CNT.
The dielectric constants of the individual segments act as scaling factors for the attractive part of the potential which is
thus varied along the CNT from segment to segment.
The only limiting factor for generating arbitrary step-wise potentials along $z$ is the availability of CNTs with the
desired electronic properties. However, nanotubes do possess versatile electronic properties and these may even be altered
by doping techniques \cite{doping14}.
\begin{figure}[t!]
\centering
\begin{overpic}[width=0.4\textwidth]{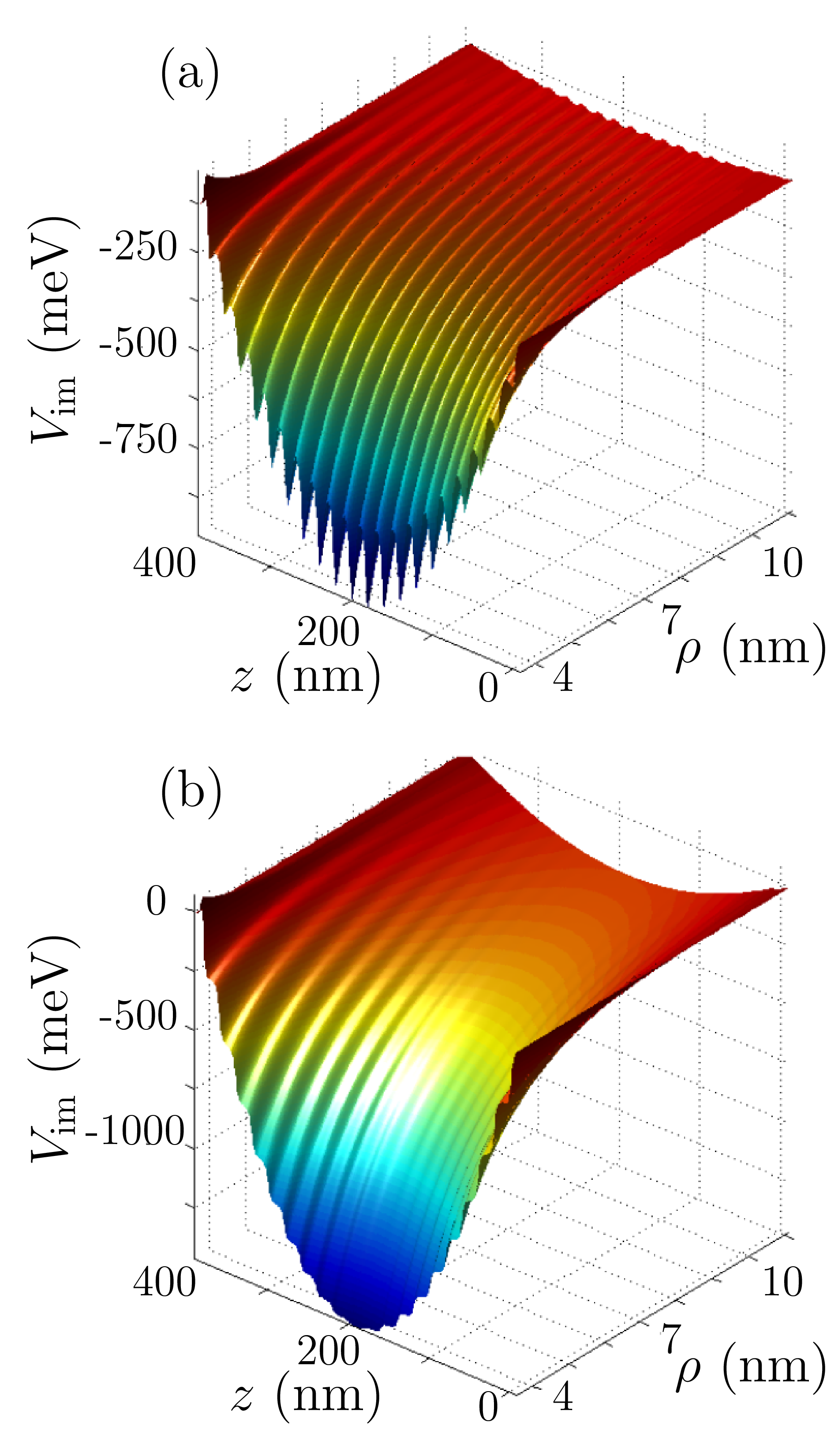}
\end{overpic}
\caption{Image potential $V_{\mbox{\tiny im}}(\rho,z)$ due to a carbon nanotube of length $L_{\mbox{\tiny CNT}} = \unit{400}{\nano\meter}$
consisting of twenty identical 20-nm-long segments fused by (a) rotationally asymmetric or (b) rotationally symmetric junctions, respectively.}
\label{fig:design}
\end{figure}
Two nanotubes can be fused by imprinting pentagon-heptagon defects into the lattice structure \cite{charlier-57,chico_defects}
which serve as junctions between both tubes.
To a certain degree, the chiral indices of the intial nanotube
segments determine the conductivity of the joint CNT which inherits the electronic properties from its constituents.
In addition, the relative rotational symmetry of the junction plays a crucial role.
The rotational symmetry of a CNT is related to the lines of allowed $\mathbf{k}$ vectors in the Brillouin zone,
and the rotational symmetry of the junction refers to how the pentagon-heptagon defects and hexagons are ordered along the
circumference of the nanotube.
Now considering a scattering process of a surface electron from one segment to another, the outcome of this strongly depends
on the relative rotational symmetries of the segments as well as on the rotational symmetry of the junction.
For example, a rotationally symmetric junction cannot mediate between two electronic surface states of different symmetries
due to energy and angular momentum conservation laws \cite{chico_symmetries}: it cannot impart any additional angular
momentum to the initial state which is then totally reflected at the junction. Therefore, depending on how two CNT segments are connected,
the whole nanotube may feature different electronic properties.

To illustrate the influence of the junctions' symmetries on the image potential of a series of joint CNT segments,
Fig.~\ref{fig:design} shows two image potentials of the same nanotubes but connected with different junctions.
Twenty segments, all $\unit{20}{\nano\meter}$ in length, are stitched together by two distinct sorts of junctions:
In the first case, segments with identical rotational symmetry properties, connected by a junction with a rotational symmetry mismatch relative to the
segments it merges, constitute a nanotube which is overall not conducting. The resulting effective potential $V_{\mbox{\tiny eff}}$
will not be smooth but feature ripples, cf.\ Fig.~\ref{fig:design}(a). In the second case, an almost smooth potential is the outcome as the
junctions possess the same symmetry as the semi-conducting segments. At fixed radial distance $\rho$, the resulting potential is
of approximately harmonic nature, cf.\ Fig.~\ref{fig:design}(b), and yields almost equidistantly spaced energy levels. The potentials were calculated by
adjusting the dielectric constants of the individual segments accordingly in Eq.~(\ref{eq:potim_finite}).

\section{Summary and outlook}\label{sec:outlook}

In summary, we have shown how multiple
excited image states can be confined and arranged along a composite nanotube, comprised of non-periodically aligned finite
nanotube segments. A framework for creating highly adjustable image potentials has been proposed, widening the range of potential control
mechanisms for these exotic states. This work paves the way for subsequent studies of many-electron systems around CNTs and their
interactions.
The Hamiltonian of a bipartite system consisting of two image states can be represented as a sum of two isolated image-state
Hamiltonians and their mutual interaction which contains the electrostatic interaction $V$, expressed
via the two-body density $\hat \rho^{\mbox{\tiny AB}}$ as

\begin{equation}\label{eq:electrostatic_interaction}
V = \int \frac{\hat \rho^{\mbox{\tiny AB}}(\mathbf{r}-\mathbf{r}^\prime)}{|\mathbf{r} - \mathbf{r}^\prime|} \mbox{d}^3 \mathbf{r} \mbox{d}^3 \mathbf{r}^\prime.
\end{equation}

\noindent Spin-dependent electron exchange and magnetic dipole-dipole interactions may be considered.
We expect strong TIS-TIS interactions at distances $< \unit{1}{\micro\meter}$.
The $R$-dependent interaction energy to second order can be computed efficiently with the wavefunctions in hand.
We plan to extend these studies to interacting TISs, spin-orbit interaction, and design of practical tube settings.
Image states on liquid helium have already been demonstrated as manipulable and strongly interacting
sets of qubits \cite{dykman_liquidhelium,platzman_liquidhelium}. Their versatility, especially the rather long lifetimes and moderate
binding energies, but also their tunability makes TIS advantageous candidates for quantum information applications.

\begin{acknowledgments}
The authors thank D. Segal for valuable comments as well as S. Kr\"onke and S. Markson for helpful
remarks on this manuscript.
J. K. and P. S. gratefully acknowledge support by the National Science Foundation through a grant for 
the Institute for Theoretical Atomic, Molecular and Optical Physics at Harvard University and Smithsonian 
Astrophysical Observatory. J. K. acknowledges financial support by the German Academic Exchange Service and
C. F. is grateful for support by the Studienstiftung des Deutschen Volkes in the framework of a scholarship.
\end{acknowledgments}


\end{document}